\documentclass[12pt]{article}

\usepackage{color}

\usepackage{graphicx}

\usepackage{eqnarray}

\usepackage{hyperref}

\usepackage{amsmath}

\usepackage{xcolor} 

\usepackage{soul} 

\begin{document}

\vskip 3mm
\baselineskip=15pt

\noindent
\begin{center}
\textbf{{\large Analysis of Complex Survival Data: a tutorial using the Shiny MSM.app application}}
\end{center}

\begin{center}
\vskip 3mm
\noindent Gustavo Soutinho$^1$ and Lu\'{i}s Meira-Machado$^2$
\vskip 2mm
\noindent $^1$EPIUnit - University of Porto\\ Rua das Taipas 135, 4050-600 Porto - Portugal, gdsoutinho@gmail.com\\

\noindent $^2$Centre of Mathematics, University of Minho, lmachado@math.uminho.pt
\noindent 
\end{center}

\vskip 8mm

\begin{center}
\end{center}

The development of applications for obtaining interpretable results in a simple and summarized manner in multi-state models is a research field with great potential, namely in terms of using open source tools that can be easily implemented in biomedical applications. In this tutorial, we introduce \texttt{MSM.app}, an interactive web application using the \texttt{Shiny} package for the \texttt{R} language. In the following sections, we present the main functionalities of the \texttt{MSM.app} and an explanation of the outputs obtained for better understanding, independent of the statistical knowledge of users.
\vskip 10mm

\noindent \textbf{1. What is the \texttt{Shiny MSM.app}?}
\vskip 4mm
The appearance of the \texttt{shiny} R package allowed to automatically share results obtained from the \texttt{R} language to be analyzed for users without any prior knowledge in terms of informatics via the internet \cite{Wojciechowski2015, Chang2017, Kaushik2016}. 

In this context, the \texttt{MSM.app} web application arises with the goal of carrying out an analysis of multi-state survival data sets. Among its functionalities, we can highlight the possibility to conduct a traditional survival analysis regarding the following topics: (i) estimation of survival functions (using the classical Kaplan-Meier estimator); (ii) comparison of survival functions between groups; and (iii) use of semi-parametric and parametric regression models to study the relationship between explanatory variables and survival time.

The \texttt{MSM.app} can also be used for the analysis of multi-state survival data that can be seen as a generalization of survival analysis in which survival is the ultimate outcome of interest but where information is available about intermediate events that individuals may experience during the study period \cite{Putter2007, MM2009, LFMM2019}. 
 
To this statistical analysis, the \texttt{Shiny MSM.app} application combines \texttt{shiny} with some other packages such as \texttt{survival} \cite{Therneau2021}, \texttt{mstate} \cite{Putter2020} and \texttt{survidm} \cite{Soutinho2021a}. 

\vskip 4mm

\noindent \textbf{2. Features of the \texttt{Shiny MSM.app}}
\vskip 4mm

The \texttt{MSM.app} is a web application that was developed to perform dynamic analysis through a set of dynamic web forms, tables, and graphics.

It is built upon two components: the user-interface scripts for the layout of the application where the outputs are displayed (\textit{ui.R}); and the other given by the server scripts with the instructions of the application (\textit{server.R}) \cite{Govan2016}.

Among the main aspects of this \texttt{Shiny} web tool that improve the flexibility and productivity, we could highlight the easy rendering of the contents without multiple reloads, the feature to add computed (or processed) outputs from R scripts, or interactively add reports and visualizations.

The \texttt{MSM.app} also provides integration with other \texttt{R} packages, javascript libraries or \texttt{CSS} customization, being under the \texttt{GPL-2} open source license. The communication between the client and server is done over the normal \texttt{TCP} connection \cite{Seal2016}. 

\vskip 4mm
\noindent \textbf{3. How is the \texttt{Shiny MSM.app} different from other applications?}
\vskip 4mm
There exists a set of available web tools specifically aimed at carrying out some parts of multi-state analysis.

Among them, we can stand out the \texttt{MSDshiny} which provides a useful and streamlined way to plan and power clinical trials with multi-state outcomes such as a view of the multi-state structure, treatment effects, or the results of different types of simulations. Another example is \texttt{MSM-shiny} application. This tool uses a \texttt{CSV} file containing multi-state data and provides the modeling and comparison of transition hazard models and the prediction of occupation probabilities \cite{Lacy2021}. Recently, \texttt{MSMplus} provides a flexible visualization of the transition probabilities, transition intensities, or probability of visiting a particular state \cite{Skourlis2021}.

After analyzing existing web applications, we have concluded that their use by non-statisticians has been limited. A possible reason for this is the lack of friendly software that covers the main goals involving survival analysis and multistate models on the same platform.

The \texttt{MSM.app} allows users to explore various types of multi-state models and perform regression inference as well as obtain several predictive measures of interest, such as the occupation probabilities, the transition probabilities, and the cumulative incidence functions. Recent methods for checking the Markov assumption are also implemented.
\vskip 4mm
\noindent \textbf{4. How to get the \texttt{Shiny MSM.app}}
\vskip 4mm
The \texttt{Shiny MSM.app} is available for free open access at the Shiny Apps repository \url{https://gsoutinho.shinyapps.io/appmsm/}.
\vskip 4mm
\noindent \textbf{5. Structure of the \texttt{Shiny MSM.app}}
\vskip 4mm

The web application consists of three parts representing different aspects of the survival analysis and its extension to complex multi-state models. 

The first one allows to perform the survival analysis from mainly of most common functions of the survival \texttt{R} packages. 

The second enables one to obtain some of the main goals of a multi-state analysis, such as the inference of regression models and the estimation of transition probabilities, through the \texttt{survidm} and \texttt{mstate} \texttt{R} packages.

Finally, \texttt{MSM.app} also includes local and global statistical tests to check the Markov assumption for multi-state using the \texttt{markovMSM} package. 

\vskip 4mm
\noindent \textbf{6. Type of input data files and requirements}
\vskip 4mm
The \texttt{MSM.app} only requires \texttt{CSV} files as input. By default, in terms of structure, the values of the data set should be separated by a comma. However, files that use a semicolon to separate the values are also accepted.

Three examples of data sets are available for consultation in the \texttt{MSM.app} from which it is possible to check the requirements that the files must have to pursue the data analysis. Each one corresponds to a specific type of structure of data: survival, illness-death model, or more general multi-state models.

\begin{figure} [h]
\centering
\includegraphics[height=4.0cm]{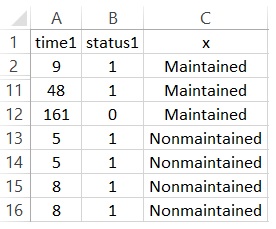}
\caption{Structure of the \texttt{CSV} files for survival analysis (given by the \textit{aml} data set).}
\label{fig:aml}
\end{figure}

The first data set corresponds to survival data in patients with acute myelogenous leukemia \cite{Miller1997}. Figure~\ref{fig:aml} shows some registers and the three columns that the file must have. ``Time1" and ``status" correspond to the time to the event and the status of the censoring, respectively. ``x" is the only covariate in this data that can be used for regression or obtaining the survival estimates for groups. 

\begin{figure} [h!]
\begin{center}
\resizebox*{5.8cm}{!}{\includegraphics{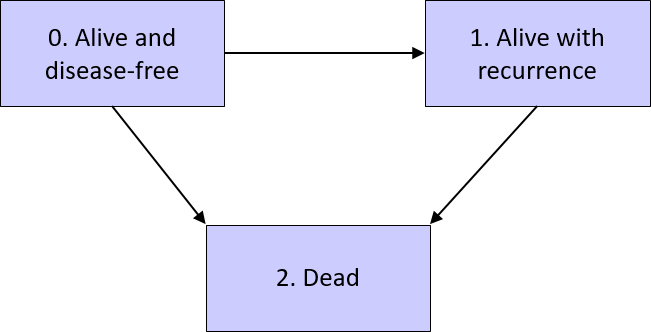}} 
\caption{Illness-death model for the colon cancer study.}
\label{fig1.1}
\end{center}
\end{figure}

\begin{figure} [h]
\centering
\includegraphics[height=4.0cm]{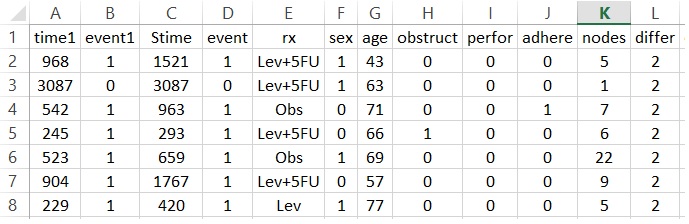}
\caption{Structure of the \texttt{CSV} files for Illness-death models analysis (given by the \textit{colonIDM} data set).}
\label{fig:colonIDM}
\end{figure}

The second corresponds to a data set from a clinical trial on colon cancer, which can be modeled using the progressive illness-death model \cite{Moertel1990}. Figure~\ref{fig1.1} shows the schematic diagram of transitions involved in the model. Among the variables of the \textit{colonIDM} data set, the first four correspond to the times or the status indicator for the illness-death model. ``time1" represents the sojourn time in the initial state and ``Stime" and ``event1" and ``event" the corresponding censoring indicates. The other variables could be used for transition probabilities or to check the effect on the transition intensities (Figure~\ref{fig:colonIDM}).

\begin{figure} [h!]
\centering
\includegraphics[width=11cm]{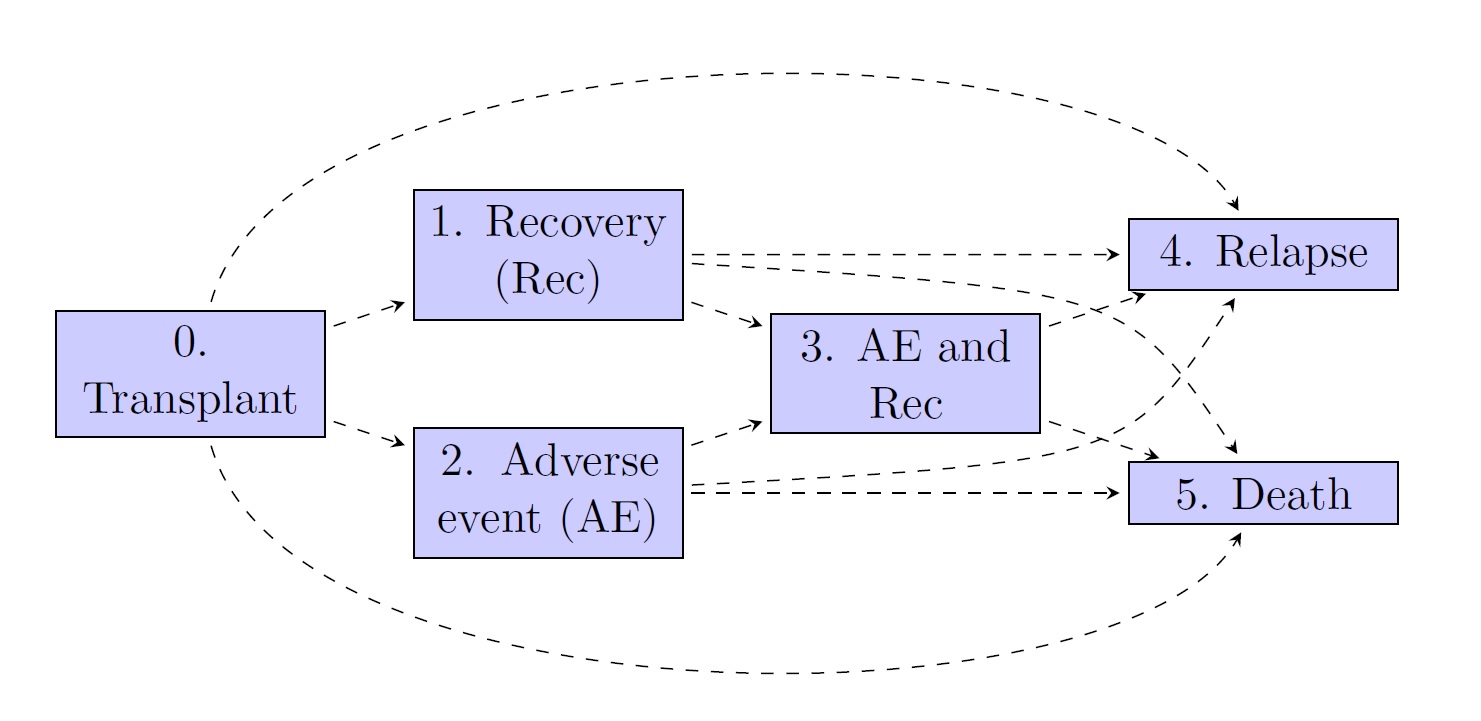}
\caption {A six-states model for leukemia patients after bone marrow transplantation.} 
\label{fig7.12}
\end{figure}

Finally, extensions to progressive processes beyond the three-state illness-death model are discussed using data from the European Group for Blood and Marrow Transplantation (EBMT) \cite{Putter2007}. The movement of the patients among the six states can be modelled through the multi-state model with the following six states: `Alive and in remission, no recovery or adverse event' (State 0); `Alive in remission, recovered from the treatment' (state 1); `Alive in remission, occurrence of the adverse event' (state 2); `Alive, both recovered and adverse event' (state 3); `Alive, in relapse' (treatment failure) (state 4) and `Dead (treatment failure)' (state 5). In total there are 12 transitions, three intermediate events given by recovery (Rec), adverse event (AE) and a combination of the two (AE and Rec), and two absorbing states: Relapse and Death  (Figure~\ref{fig7.12}).

\begin{figure} [h]
\centering
\includegraphics[height=4.0cm]{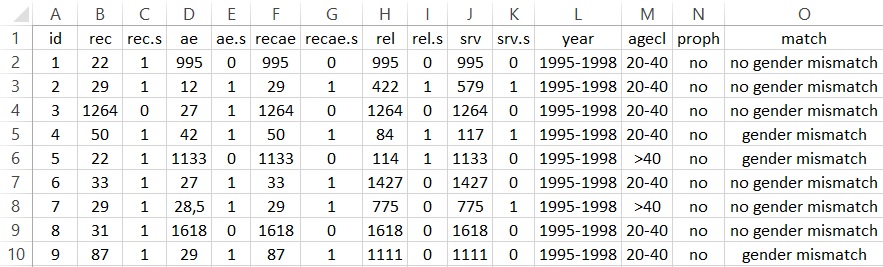}
\caption{Structure of the \texttt{CSV} files for multi-state models analysis (given by the \textit{ebmt4} data set).}
\label{fig:ebmt4}
\end{figure}

In terms of format, the \texttt{CSV} file should be in wide format. In the case of the \textit{ebmt4} data (Figure~\ref{fig:ebmt4}), the transition times from the initial to the ultimate state (`srv’) or to the intermediate states (`rec’, `ae’, `recae’ and `rel’), and the corresponding censoring variables are rec.s’, `ae.s’, `recae.s’, `rel.s’ and `srv.s’. Other variables correspond to the covariates for regression models.

For all these data sets, the names of the variables could be different. In these cases, the user must correctly indicate the corresponding name in the web forms of the application. The data files can be accessed at \url{https://w3.math.uminho.pt/~lmachado/shiny/}.

\vskip 4mm

\noindent \textbf{7. How to select the input files?}
\vskip 4mm

\begin{figure} [h]
\centering
\includegraphics[width=4.4cm, height=4.4cm]{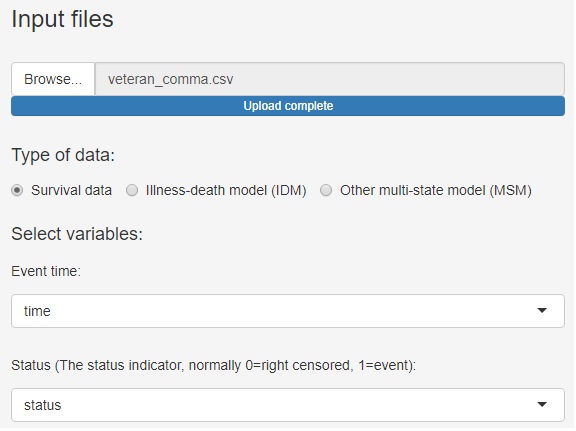}
\includegraphics[width=4.4cm,height=4.6cm]{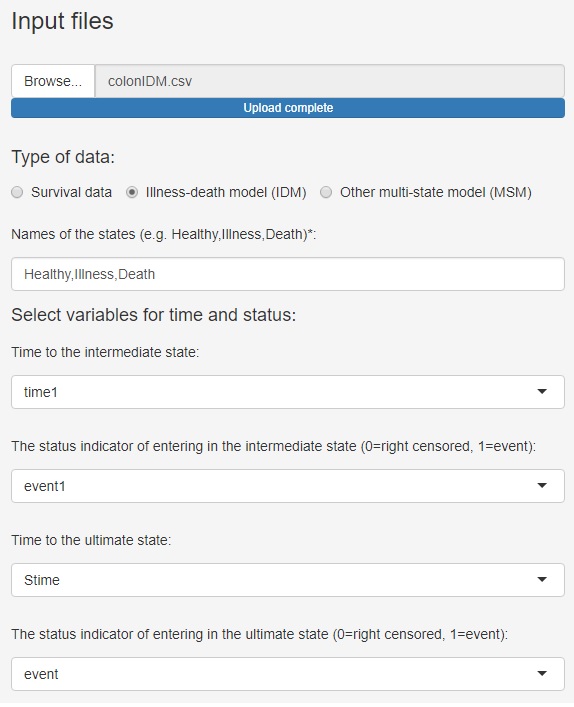}
\includegraphics[width=4.4cm,height=4.6cm]{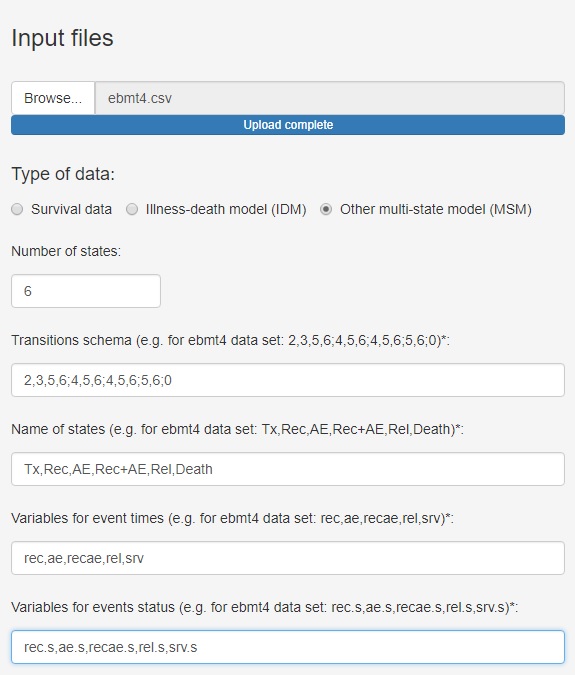}
\caption{The event time and status variables for the survival analysis from the \textit{veteran} data set (left); an indication of the two event times and their corresponding status for the illness-death model given by the \textit{colonIDM} data (center); and a description of the MSM model through the number and the state names, the transition schema, and the event times and corresponding status for the \textit{ebmt4} data (right).}
\label{fig:input2}
\end{figure}

On the ``input files" page we can find an interactive form. For each type of data set (``survival data", ``illness-death model", or ``multi-state model"), a new web form appears below the radio buttons, in which we indicate the times to the events or the status indicator (Figure~\ref{fig:input2}, left hand side and center). In case of the more complex multi-state models, it is also necessary to indicate the transition schema as well as the number of states, for instance (Figure~\ref{fig:input2}, right hand side).

\begin{figure} [h!]
\centering
\includegraphics[height=3.0cm]{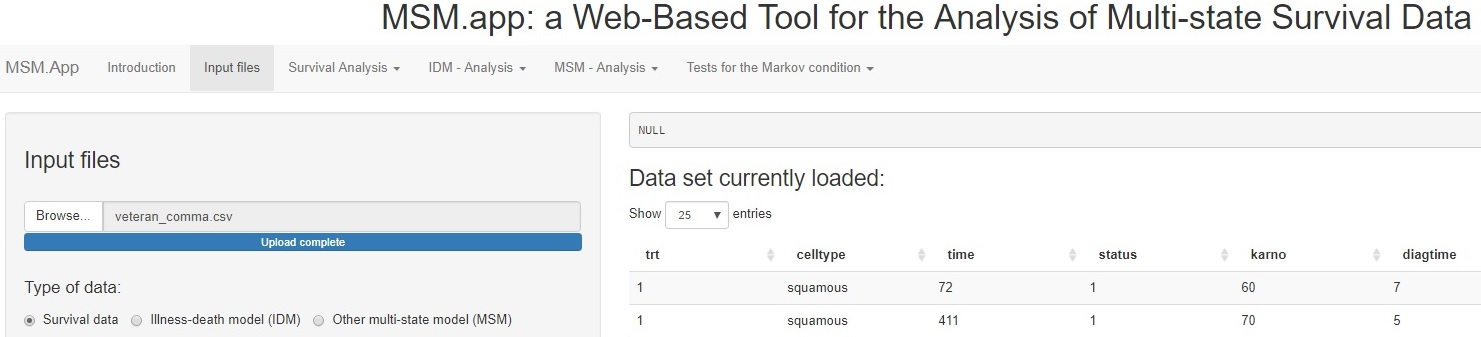}
\caption{The input file page with the data table with some results for the \textit{veteran} data set and three radio buttons for each type of model.}
\label{fig:input}
\end{figure}

After submitting a data set, a table appears on the right hand side of the page which can be dynamically. This table can be changed using filters or by searching for specific words in the table. Figure~\ref{fig:input} shows a partial view the \textit{veteran} data set that can be found in the \texttt{survival} package.

\vskip 10mm
\noindent \textbf{8. How to carry out a Survival analysis?}
\vskip 4mm
From the ``survival analysis" button, we can carry out the classical methods for survival analysis. 
In this section, we analyze the main aspects of the outputs displayed on the right sides of the pages: Kaplan-Meier estimator, Compare survival curves, Cox PH models and parametric models, by using the \textit{veteran} data set. 

\vskip 2mm
\textit{Kaplan-Meier estimator}
\vskip 2mm

\begin{figure} [h]
\centering
\includegraphics[height=4.8cm]{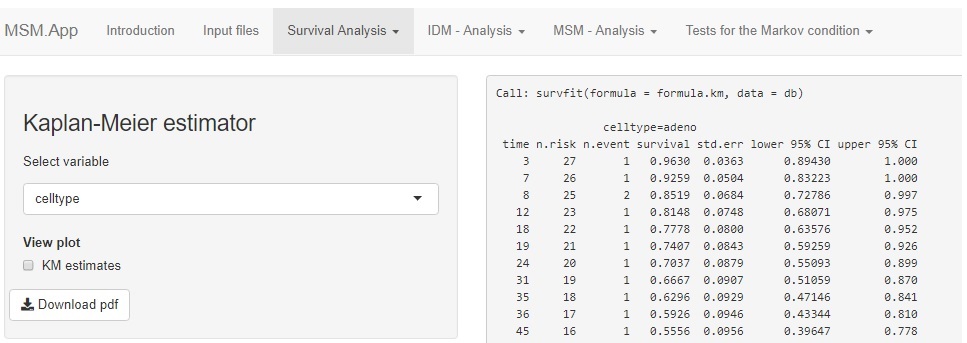}
\caption{Summary of the survival estimation for the categorical covariate ``celltype" of the \textit{veteran} data set using the Kaplan-Meier estimator.}
\label{fig:km_summary}
\end{figure}

\begin{figure} [h]
\centering
\includegraphics[height=5.8cm]{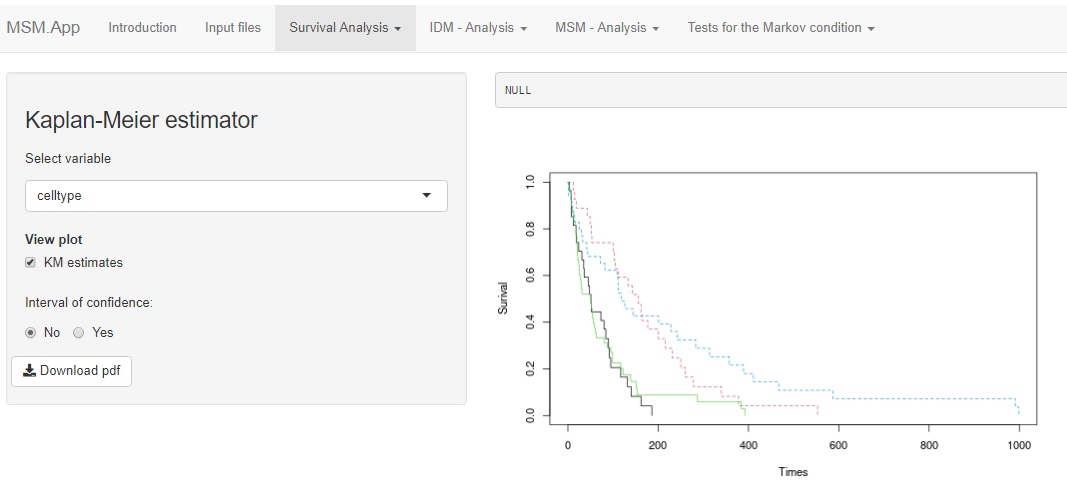}
\caption{The output of the survival estimation for the categorical covariate ``celltype" of the \textit{veteran} data set using the Kaplan-Meier estimator. Survival curves for each group with confidence intervals could be shown at the bottom by choosing ``Yes".}
\label{fig:km}
\end{figure}

Non-parametric estimation of the survival function is traditionally performed using the Kaplan-Meier estimator, which can be desegregated for different groups of categorical variables. By default, a summary of the estimates is presented on the right side. Figure~\ref{fig:km_summary} shows, respectively, times with events, the number of individuals at risk at that time, and the survival estimates with corresponding standard errors and confidence intervals. It is also possible to display an interactive plot of the survivals with (or without) the confidence interval intervals (Figure~\ref{fig:km}).

\vskip 2mm
\textit{Compare survival curves}
\vskip 2mm

\begin{figure} [h]
\centering
\includegraphics[height=4.2cm]{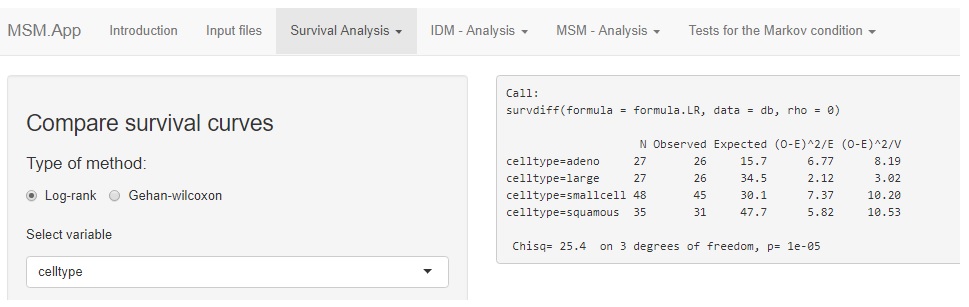}
\caption{The output of the Log-rank test for the categorical covariate ``celltype" of the \textit{veteran} data set for testing the null hypothesis of no difference in survival between the two groups.}
\label{fig:logrank}
\end{figure}

Statistical tests can be used to compare survival rates between groups. The null hypothesis states that there is no difference in survival between groups. The log-rank test and the Gehan-Wilcoxon test are the most commonly used. Both tests are available in the \texttt{MSM.App}. The output of the tests comprises the chi-squared statistics, degrees of freedom, and the corresponding p-value. The number of individuals for each group and the number of observed and expected values are also presented. Results of Figure~\ref{fig:logrank} show significant differences in the survival curves among groups given by the cell types ($p<$0.0001).

\begin{figure} [h]
\centering
\includegraphics[height=5.8cm]{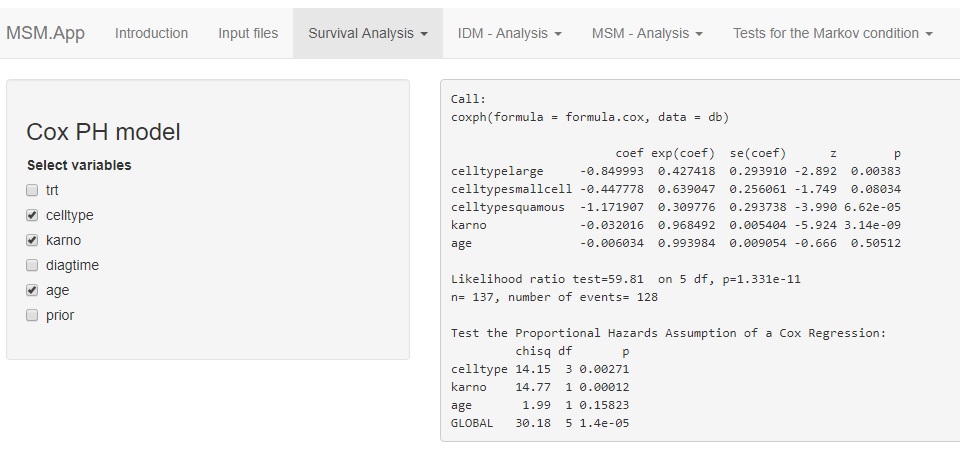}
\caption{Results of the Cox model for the variables ``celltype", ``karno" and ``age" of the \textit{veteran} data set. Result of the test for the proportional hazard assumption shows that a Cox PH model  }
\label{fig:coxph}
\end{figure}

\vskip 2mm
\textit{Cox PH models}
\vskip 2mm
The semi-parametric Cox proportional hazards model \cite{Cox1972} is usually used to evaluate the effects of several factors on survival. The output for the Cox model is updated as the user selects the variables to be included in the model in the web form. Figure~\ref{fig:coxph} shows, respectively, the estimate of the coefficients for the covariates ``celltype", Karnofsky performance score (``karno") and ``age";  the hazard ratio given by $exp(coef)$ and the respective standard error; and the statistical significance of the model. Result of the global test for the proportional hazards assumption confirms that this requirement is no fulfilled and consequently a parametric accelerated failure time (AFT) model must be preferable to be fitted to this survival data.

\begin{figure} [h]
\centering
\includegraphics[height=5.4cm]{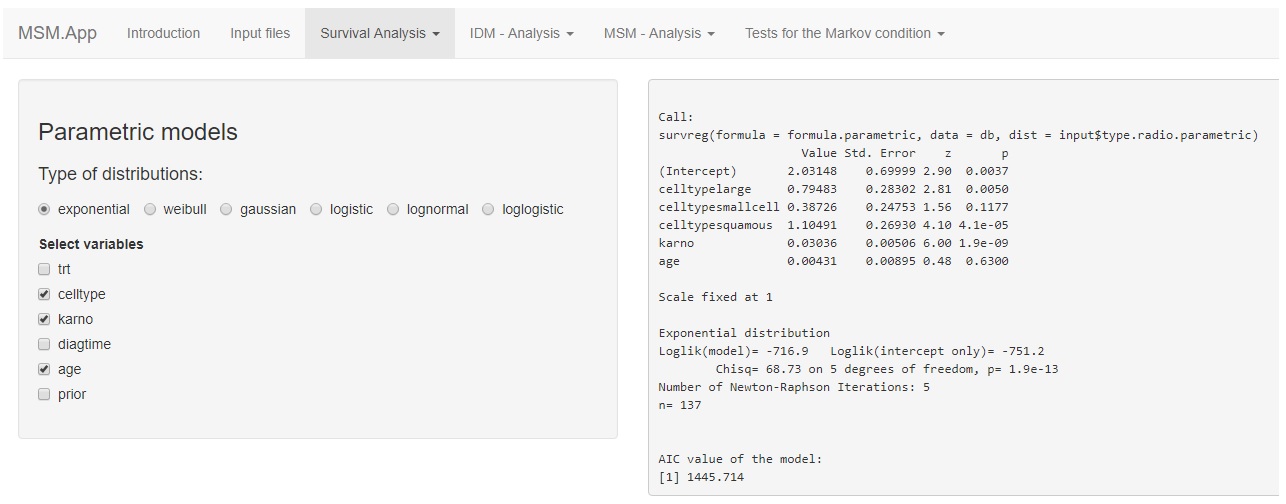}
\caption{The outputs of the parametric models page with the results of the fitted model for the variables ``celltype", ``karno" and ``age" of the \textit{veteran} data set using the exponential distribution.}
\label{fig:aft_exp}
\end{figure}

\begin{figure} [h]
\centering
\includegraphics[height=5.4cm]{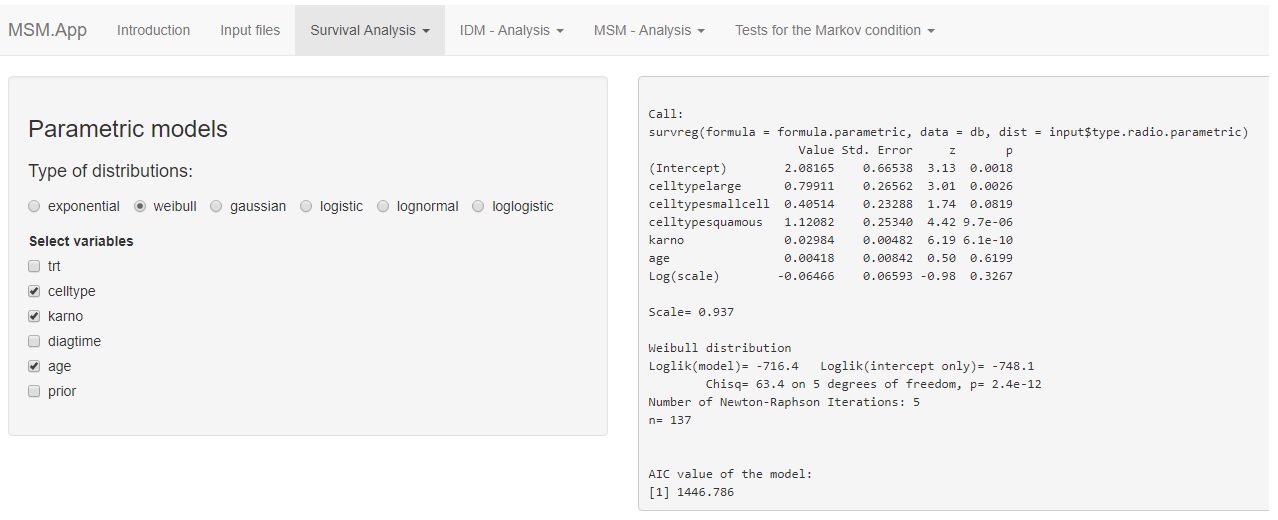}
\caption{The outputs of the parametric models page with the results of the fitted model for the variables ``celltype", ``karno" and ``age" of the \textit{veteran} data set using the weibull distribution.}
\label{fig:aft_wei}
\end{figure}

\begin{figure} [h]
\centering
\includegraphics[height=5.4cm]{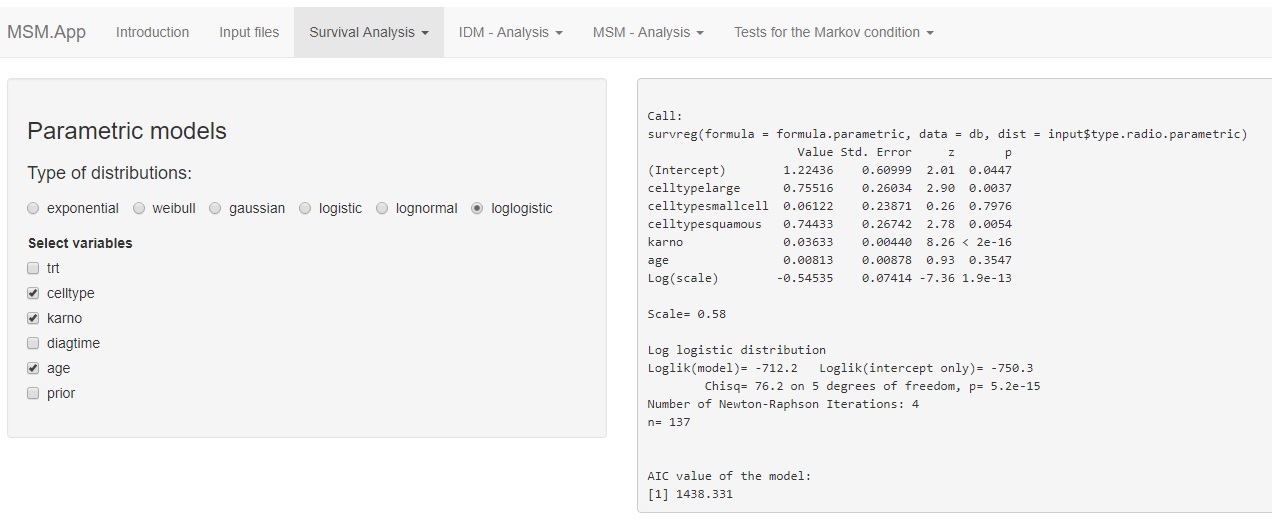}
\caption{The outputs of the parametric models page with the results of the fitted model for the variables ``celltype", ``karno" and ``age" of the \textit{veteran} data set using the loglogistic distribution.}
\label{fig:aft_log}
\end{figure}

\vskip 2mm
\textit{Parametric AFT models}
\vskip 2mm
The idea behind the outputs for parametric survival models is quite similar to those provided by the Cox PH models. In this case, six possibilities of distributions are available to model the baseline hazard function of the models: exponential, weibull, gaussian, logistic, lognormal or loglogistic. Besides the summary of the models, the Akaike Information Criterion (AIC) values are also presented to make it easier to compare which model fits better to assess the effect of several risk factors on survival time. Figures~\ref{fig:aft_exp}, \ref{fig:aft_wei} and \ref{fig:aft_log} show, respectively, the outputs of the AFT models given by the exponential, weibull and loglogistic distributions. Through the AUC values we can assume that loglogistic parametric model is preferable. From this model, we can observe that only ``age” and ``small cells” do not have a significant effect on survival.  

\vskip 2mm
\noindent \textbf{9. How to carry out a illness-death model Analysis?}
\vskip 2mm

\begin{figure} [h]
\centering
\includegraphics[height=2.4cm]{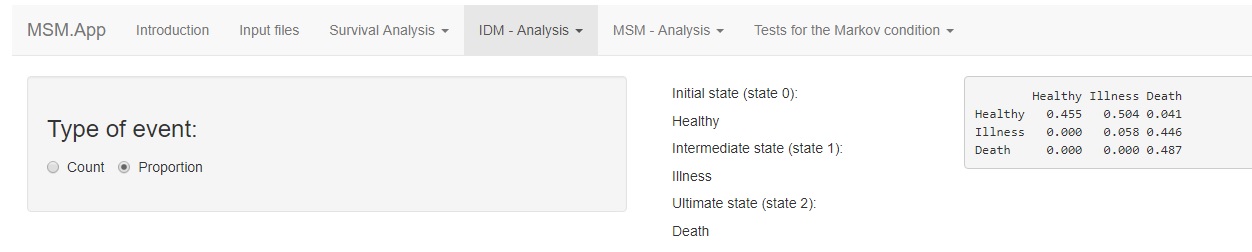}
\caption{The number of transitions among the three states of the \textit{colonIDM} data set.}
\label{fig:ntrans}
\end{figure}

From the `` IDM-Analysis" button, we can carry out a data analysis of a progressive illness-death (IDM) model given by two events, three states, and three transitions. 
\vskip 2mm
\textit{Number of events}
\vskip 2mm
First, we are interested in having an idea of the movement of individuals among the three states. It is also possible to see the proportions of transitions (Figure~\ref{fig:ntrans}).

\begin{figure} [h]
\centering
\includegraphics[height=6cm]{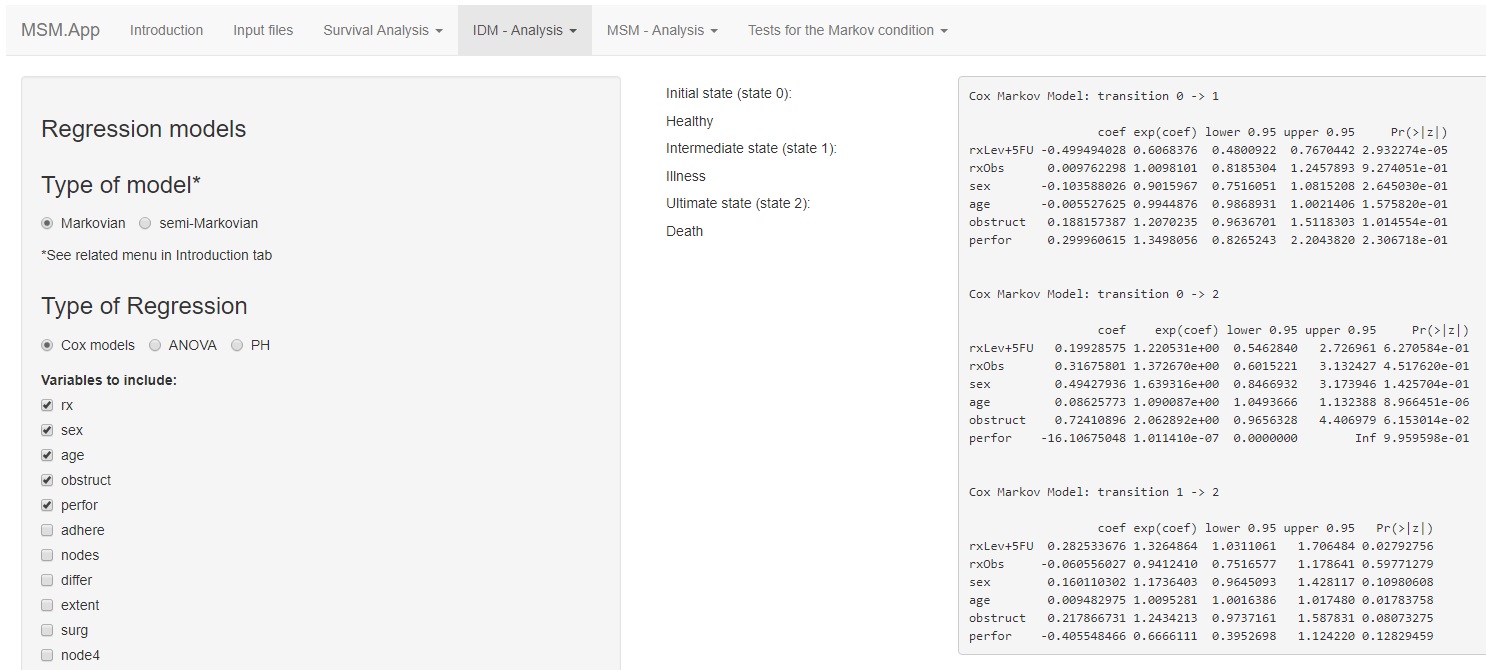}
\caption{Results of the application of the Cox PH model with the following covariates: ``rx", ``sex", ``age", ``obstruct" and ``perfor". Results for each of the three transition intensities of the \textit{colonIDM}. A Markovian process is assumed.}
\label{fig:regidm}
\end{figure}

\vskip 2mm
\textit{Regression models}
\vskip 2mm
In the multi-state models, there are so many transition intensities as there are transitions. For each one, we can then check the effect of the individual characteristics of the individuals by fitting separate intensities using semi-parametric Cox proportional hazard regression models \cite{Cox1972}. In the inference of the regression models, we must take into account if the Markov condition, in which the past and future are independent given the present state, is verified. Regarding the dependence of the transition intensities and time, in the case of the failure of the markovianity, we can use a semi-Markov model in which the future of the process does not depend on the current time but rather on the duration of the current state. These two methods are available in \texttt{MSM.app}. In terms of interpretation, the outputs of the regression, for each transition, are very similar to those presented for the survival analysis. Results of Figures~\ref{fig:regidm} indicate that save the treatment ``Lev(amisole)+5-FU” all the other covariates have no effect for recurrence transition and only age could be considered important on mortality without recurrence. At least, only the covariates ``perfor”, ``sex” and the treatment ``Obs” have no association for the transition from recurrence to mortality.

\begin{figure} [h]
\centering
\includegraphics[height=6.2cm]{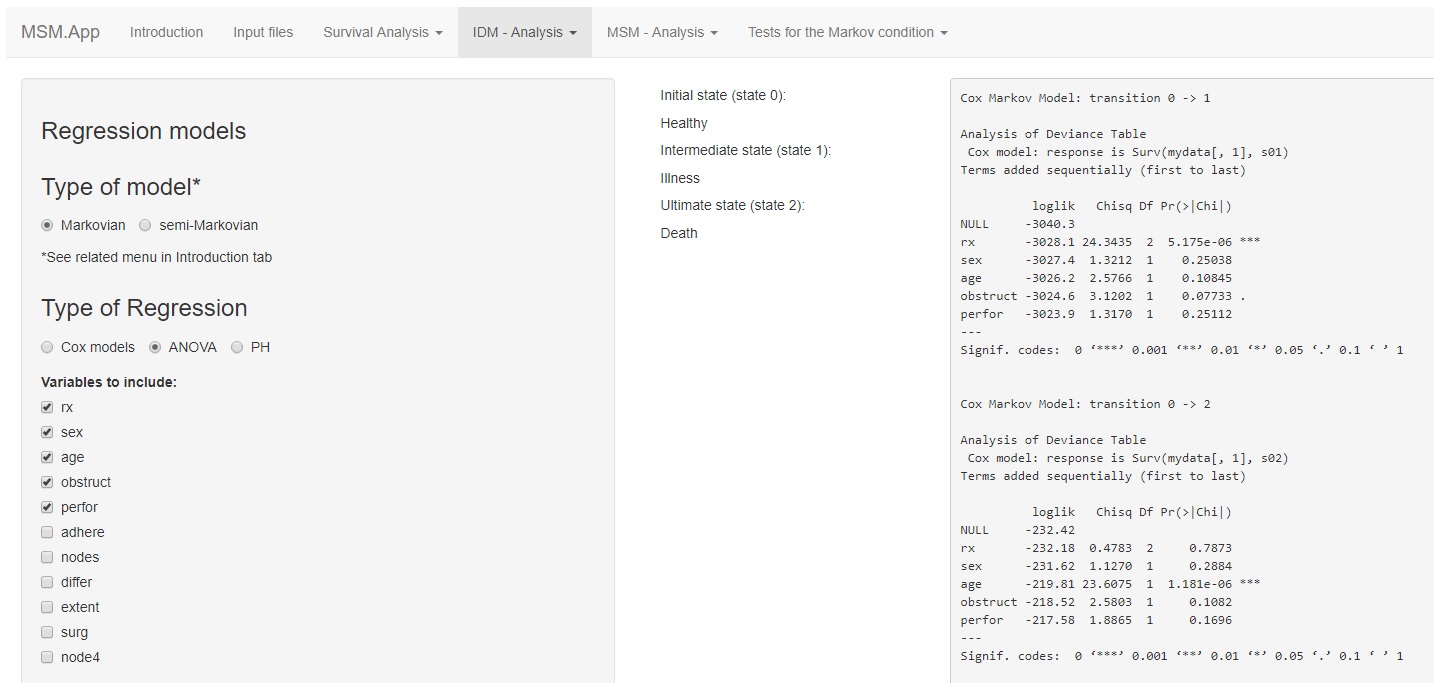}
\caption{ANOVA results for transitions $0\longrightarrow1$ and $0\longrightarrow2$ with the following covariates: ``rx", ``sex", ``age", ``obstruct" and ``perfor" of the \textit{colonIDM}. A Markovian process is assumed.}
\label{fig:regidm-anova}
\end{figure}

\begin{figure} [h]
\centering
\includegraphics[height=4.5cm]{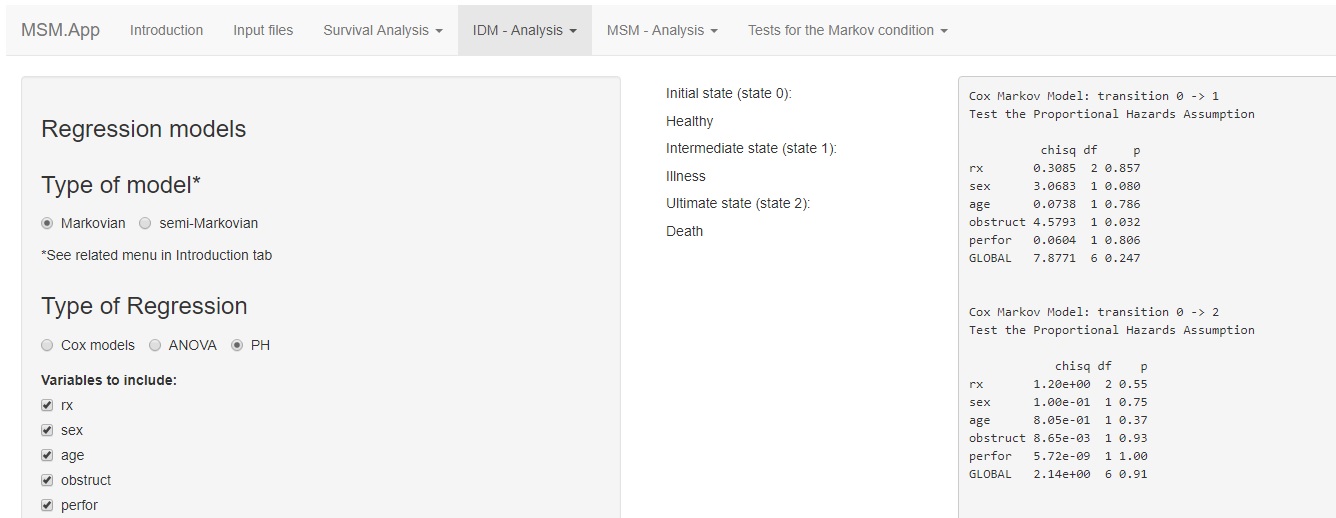}
\caption{The proportional hazards assumption was tested for the transitions $0\rightarrow1$ and $0\rightarrow2$ with the following covariates: ``rx", ``sex", ``age", ``obstruct" and ``perfor" of the \textit{colonIDM}. A Markovian process is assumed.}
\label{fig:regidm-ph}
\end{figure}

It is also possible to obtain the outputs of ANOVA tests and the p-values of the tests for nonlinearity. For both outputs, the summaries show the chi-squared statistics and the p-values. For ANOVA tests, the log-likelihood values for each parameter are also presented (Figures~\ref{fig:regidm-anova} and~\ref{fig:regidm-ph}).

\begin{figure} [h]
\centering
\includegraphics[height=7.1cm]{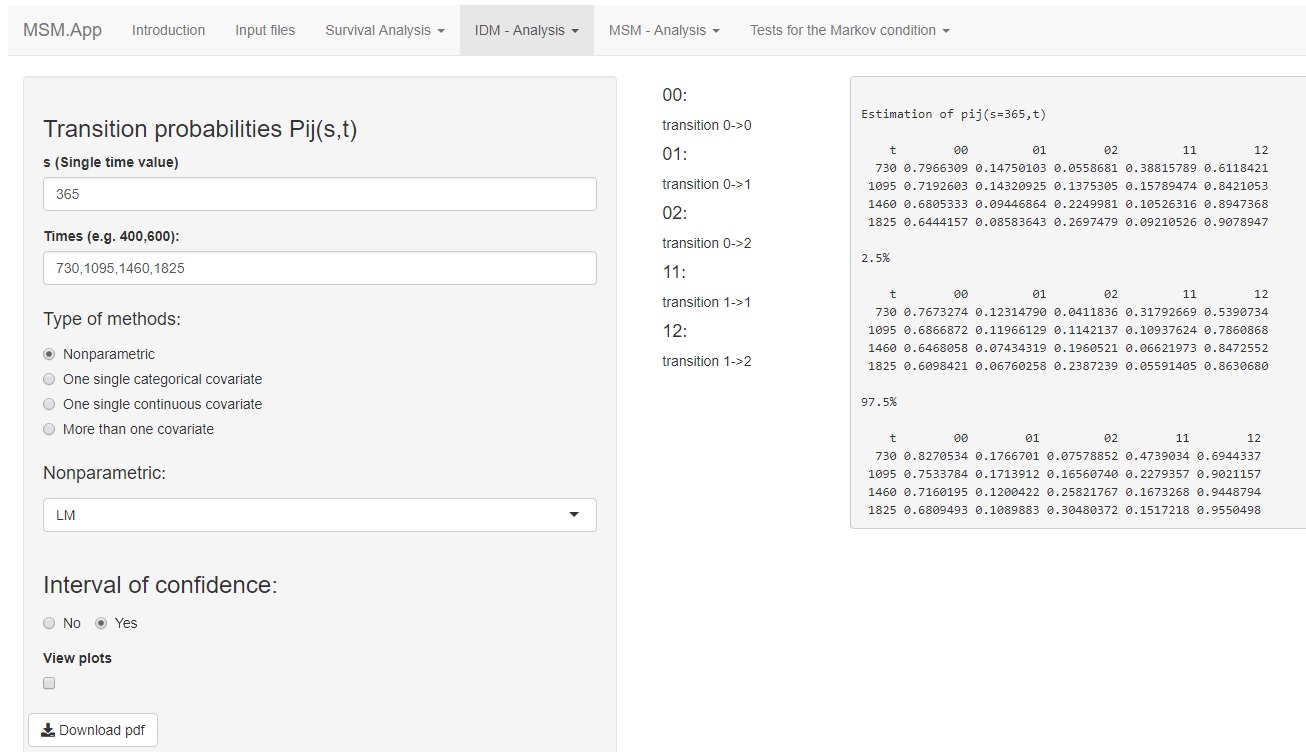}
\caption{Results of the estimates of the transition probabilities and the corresponding confidence intervals for \textit{s} = 365 and times 730, 1090, 1460, and 1825 days for the \textit{colonIDM} data set using the landmark estimators.}
\label{fig:tplm}
\end{figure}

\begin{figure} [h]
\centering
\includegraphics[height=7.6cm]{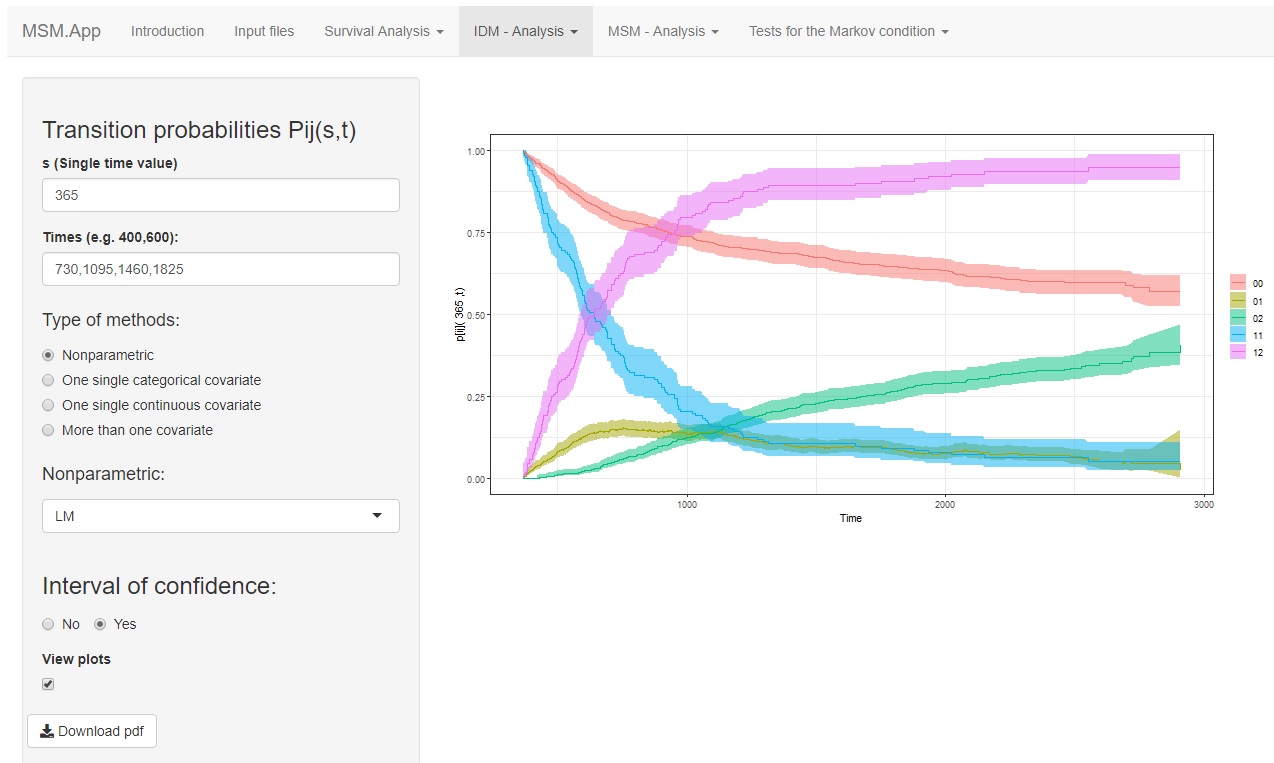}
\caption{Transition probability estimates with confidence intervals for each transition for \textit{s} = 365 using the landmark estimators.}
\label{fig:tplm-plot}
\end{figure}

\begin{figure} [h]
\centering
\includegraphics[height=8.2cm]{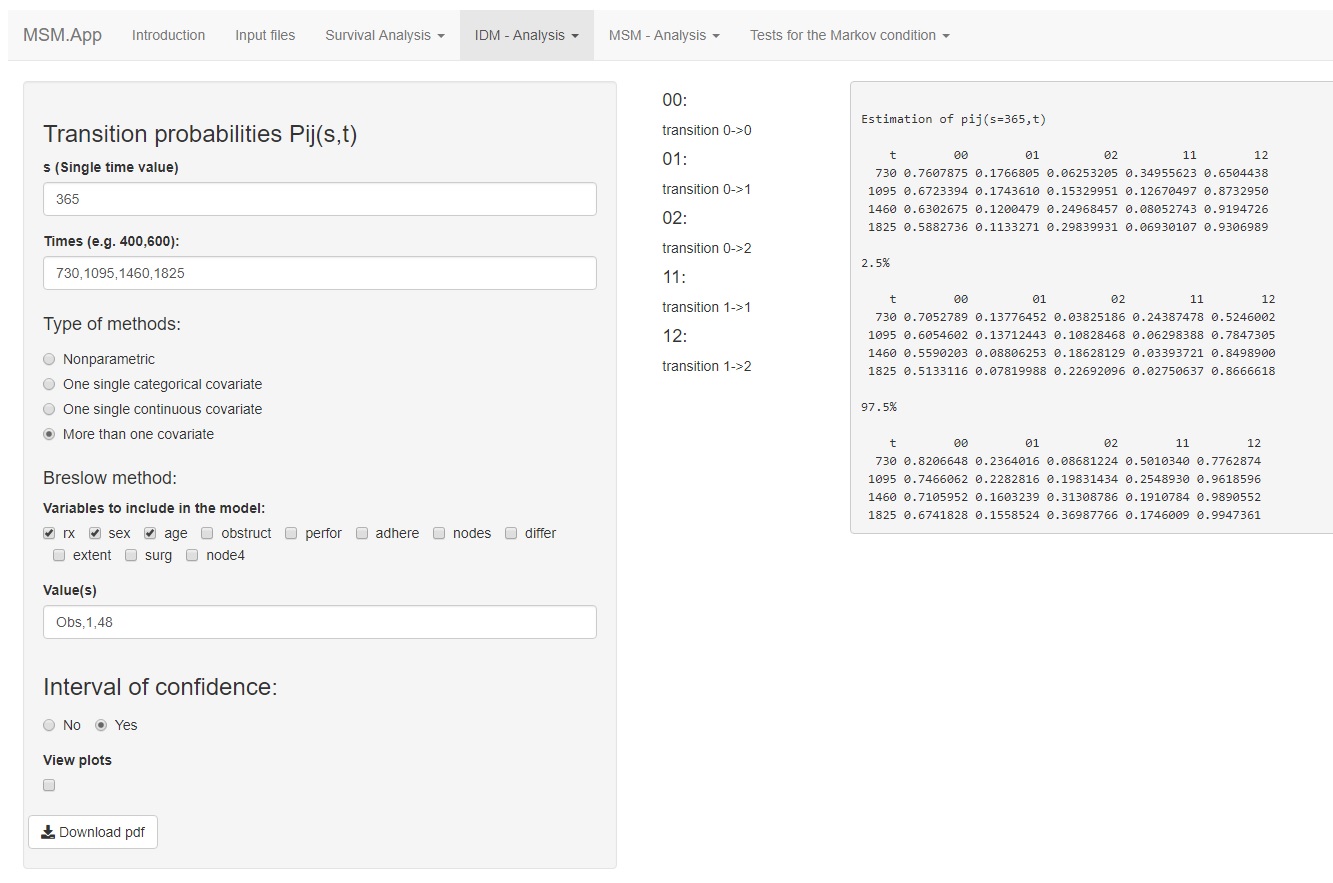}
\caption{Results of the estimates of the transition probabilities and the corresponding confidence intervals for \textit{s} = 365 and times 730, 1090, 1460, and 1825 days for the \textit{colonIDM} data set using the Breslow estimator.}
\label{fig:tpBreslow}
\end{figure}

\vskip 2mm
\textit{Transition probabilities}
\vskip 2mm
The transition probabilities are quantities of particularly interest since they allow for long-term predictions of the multi-state process. The \texttt{MSM.app} allows estimating these quantities using the Aalen-Johansen estimator \cite{Aalen1978}, the landmark methods (\texttt{LM}) proposed by \cite{AlvarezMM2015}, as well as its presmoothed version (\texttt{PLM}) proposed by \cite{Machado2016}, and the landmark Aalen-Johansen \cite{PutterSpitoni2018}. Related references can also be seen in the papers by \cite{Moreira2013} and \cite{AA2014}. Categorical covariates can be included using all four of these methods by splitting the sample for each level of the covariate and repeating the described procedures for each subsample. Through the \texttt{IPCW} estimator proposed by \cite{MM2015} is also possible to estimate transition probabilities conditional on one single continuous covariate. Finally, the \texttt{MSM.app} also provides the estimation of transition probabilities conditional on several covariates through Breslow's method for estimating the baseline hazard function of the Cox models fitted marginally to each transition. The outputs for all these methods are identical. As an example, Figure~\ref{fig:tplm} shows the estimates of the transition probabilities, using the landmark approach, from the initial single time $s = 365$ days to the next four years (730, 1095, 1460, and 1825 days). Results are presented combining the values of the corresponding times and transitions which are labeled from ``00" to ``12".  For instance, ``01" corresponds to the transition from the initial state (State 0) to intermediate state (Recurrence, State 1) and, in similar way, ``12” the transition from the intermediate state (Recurrence, State 1) to death (State 2). Plots for each five transition probabilities are shown in Figure~\ref{fig:tplm-plot}. $\widehat p_{00}(s=365,t)$ corresponds to the probability of a individual to occupy the initial state at time $t$ conditional to be in the same state at time $s=365$. In a similar way, $\widehat p_{11}(s,t)$ represents the conditional probability of those individuals observed in State 1 at time $s=365$ to remain in the intermediate state at a later time $t$. Plots for these transition probabilities report, respectively, survival fractions along times among the individuals that belong to initial state and the intermediate state at time $s=365$, being represented by monotone non-increasing functions. Plots for $\widehat p_{02}(s,t)$, report one minus the survival fraction along time, among the individuals in the initial state at time $s$. In this case, the plot is given by a monotone non-decreasing function. Finally, plots for $\widehat p_{01}(s,t)$ allows for an inspection along time of the probability of being in State 1 for the individuals who belong to State 0 at time $s$. As expected the confidence bands become wider with greater lags times $t-s$. Finally, Figure~\ref{fig:tpBreslow} depict the estimates of the transition probabilities for individuals with covariate values (``rx",``sex", and ``age") as \texttt{Obs,1,48}, respectively. 

\begin{figure} [h]
\centering
\includegraphics[height=7.8cm]{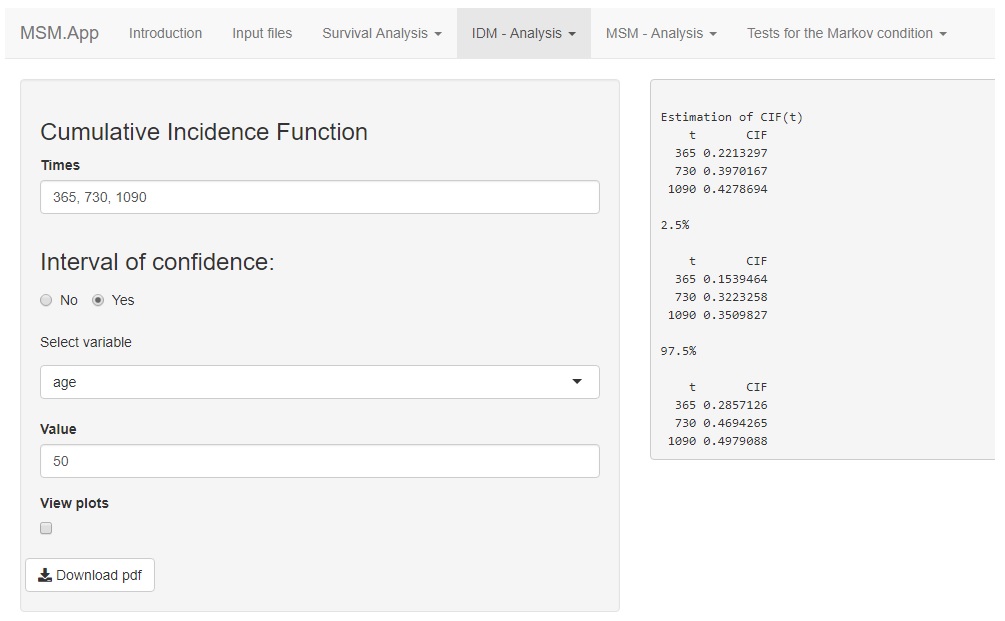}
\caption{Cumulative recurrence incidence with 95\% bootstrap confidence intervals. Data from a colon cancer study.}
\label{fig:cif}
\end{figure}

\vskip 2mm
\textit{Cumulative Incidence Function (CIF)}
\vskip 2mm
The cumulative incidence of the illness (intermediate state) is another quantity of interest in IDM models \cite{kalbfleisch}. This quantity denotes the probability of the individual or item being or having been in the intermediate ‘diseased’ state at some particular time $t$. It can be estimated conditional on a covariate, continuous or categorical. Figure~\ref{fig:cif} shows the estimates, and respective bounds of confidence, of the CIF conditional to ``age” at 50 years for three specific times.

\vskip 4mm
\noindent \textbf{10. How to carry out a multi-state analysis for other models?}
\vskip 2mm

\begin{figure} 
\centering
\includegraphics[width=12.5cm]{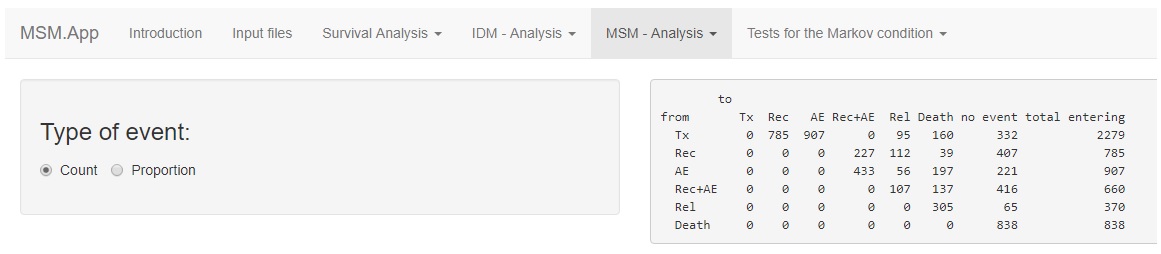}
\caption{The number of transitions among the states of the \textit{ebmt4} data set.}
\label{fig:ntrans.msm}
\end{figure}

\begin{figure} 
\centering
\includegraphics[width=12.5cm]{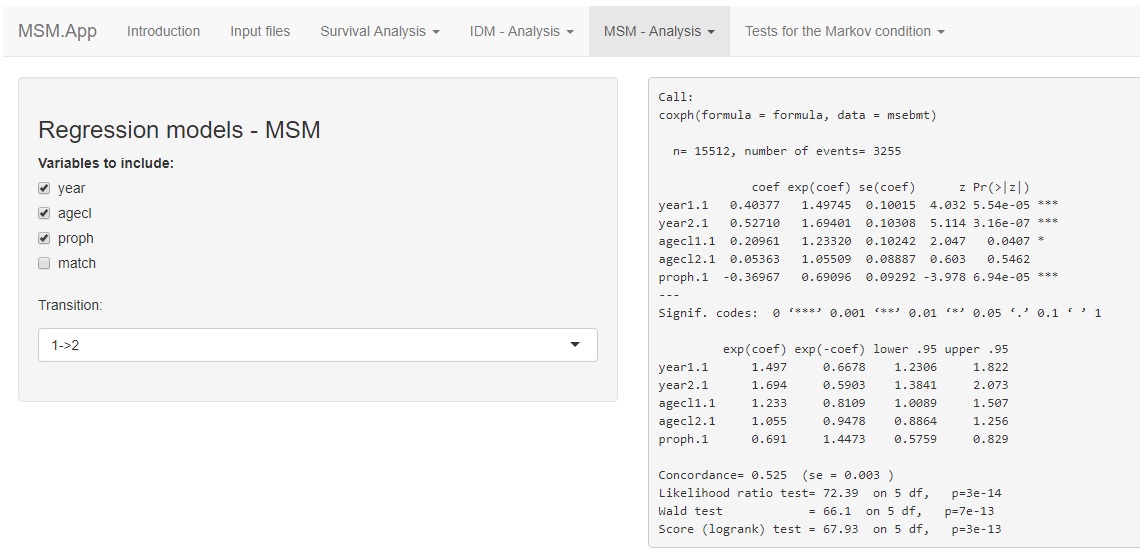}
\caption{The output of the Cox regression model for the transition $1\rightarrow2$ that include the `year", ``age" and ``proph" covariates. \textit{ebmt4} data set.}
\label{fig:msm-reg}
\end{figure}

\begin{figure} 
\centering
\includegraphics[height=6.7cm]{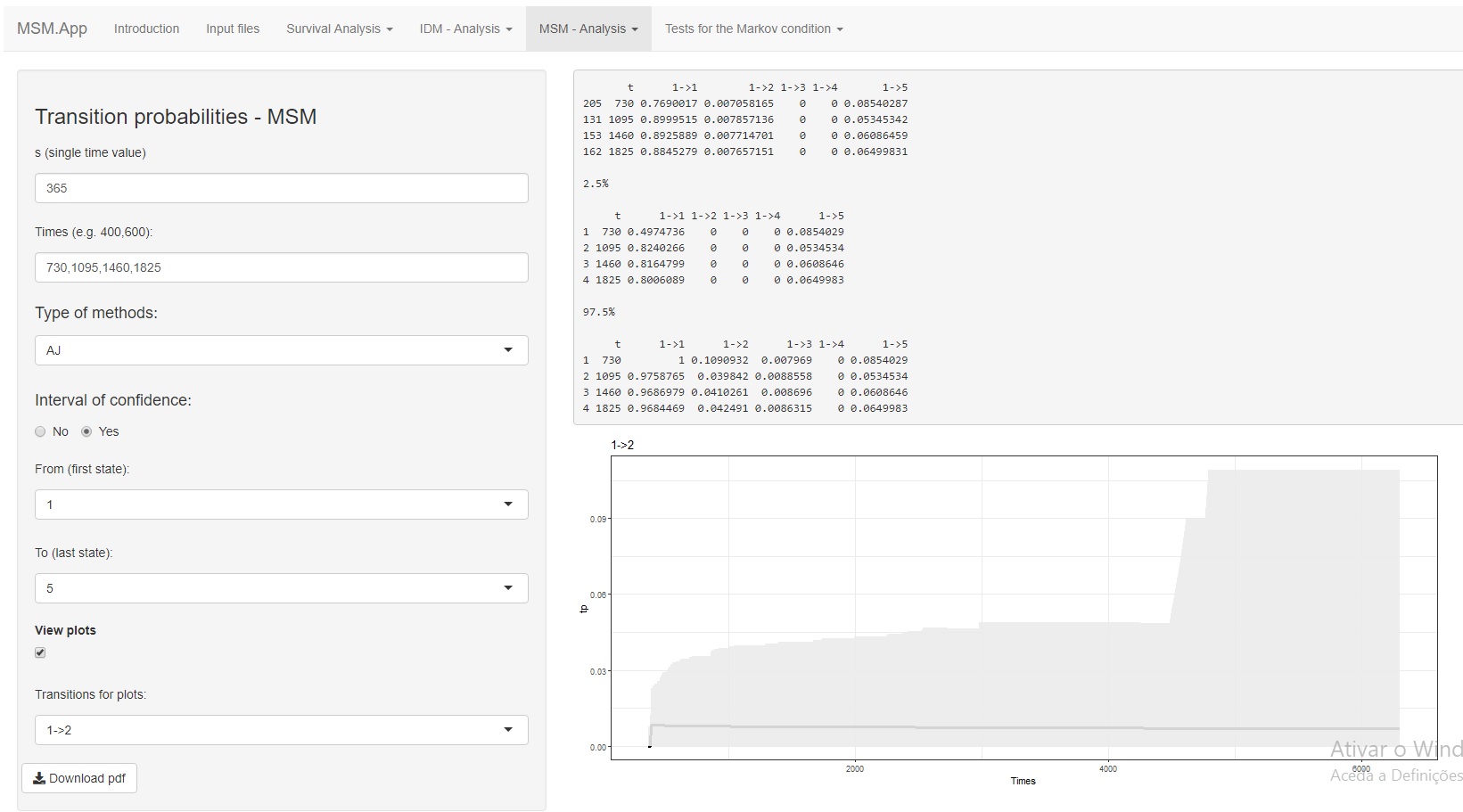}
\caption{Estimates of all possible transition probabilities from the state 1 to 5, for \textit{s} = 365 and times equal to 730, 1095, 1460, and 1825 using the \texttt{AJ} estimators. \textit{ebmt4} data set.}
\label{fig:tpmsm}
\end{figure}

By clicking on the ``MSM-analysis", we can extend some of the methods addressed for the IDM models to more complex multi-state models (MSM) with more than three states and possible reversible transitions.
\vskip 2mm
\textit{Number of events}
\vskip 2mm
Figure~\ref{fig:ntrans.msm} shows the movement of the individuals among the six states of the multi-state model represented by the data set of the European Group for Blood and Marrow Transplantation (\textit{ebmt4} data set).
\vskip 2mm
\textit{Regression models}
\vskip 2mm
Figure~\ref{fig:msm-reg} shows the results of the Cox regression model for the transition $1\rightarrow2$ which includes the covariates ``year", ``age" and ``proph". As we can see, in terms of interpretation, the output is quite similar to those of IDM models but also shows the global significance of the model through different tests.
\vskip 2mm
\textit{Transition probabilities}
\vskip 2mm
The steps for obtaining the transition probabilities are identical to those used for the illness-death model. In Figure~\ref{fig:tpmsm}, the output shows all the estimates from the indicated start and last states of the transition probabilities as well as the corresponding confidence intervals. A plot with the transition probability can also be presented for each transition.
 
\vskip 4mm
\noindent \textbf{11. How to check the Markov condition?}
\vskip 4mm

Traditionally the Markov assumption is checked by including covariates depending on the history through a proportional hazards model. Since the landmark methods of the transition probabilities are free of the Markov assumption, they can also be used to introduce such tests by measuring their discrepancy to Markovian (\texttt{AJ}) estimators.

The \texttt{MSM.app} web application offers two types of tests for checking this assumption using recent literature methods: (i) local tests, which are obtained by fixing a specific time value, $s$ and are particularly useful for estimating transition probabilities; and (ii) global tests, which may be preferable for regression purposes.

\vskip 2mm
\textit{Local tests}
\vskip 2mm

\begin{figure} [h]
\centering
\includegraphics[height=9.9cm]{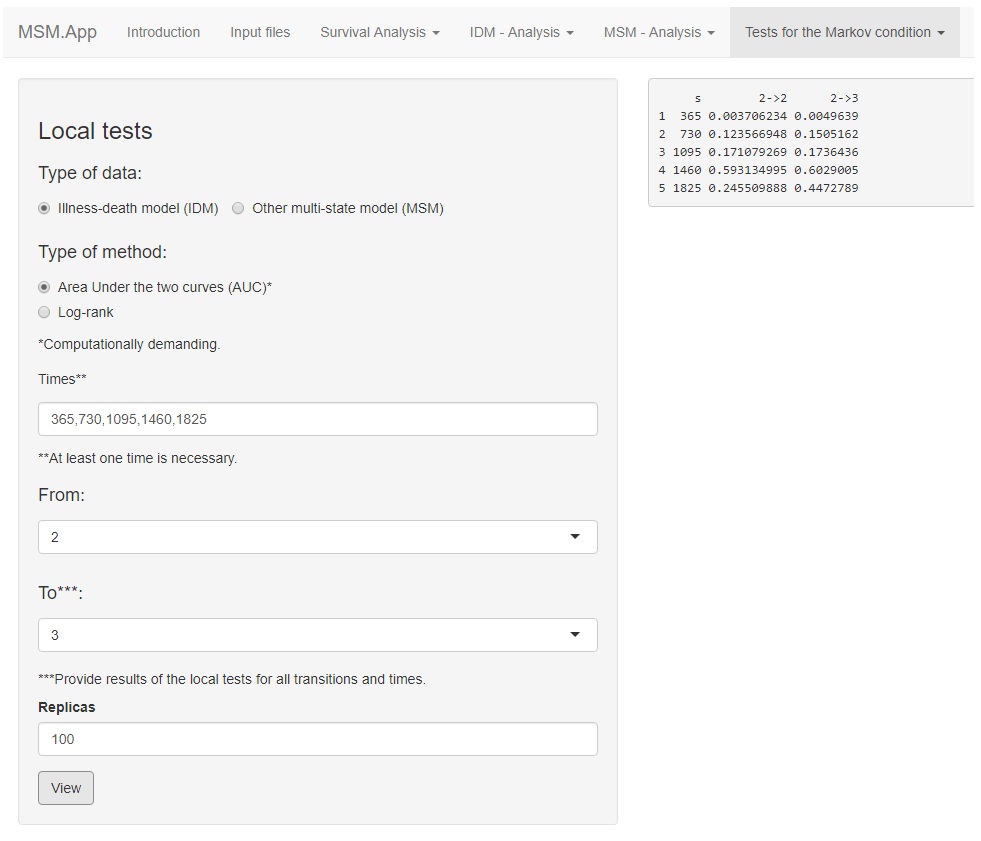}
\caption{Results of the local test for the illness-death model using the colon cancer data set, for \textit{s} = 365, 730, 1095, 1460, and 1825 days, using the \texttt{AUC} test, from state 1 to state 3.}
\label{fig:markov-localTest}
\end{figure}

\begin{figure} [h]
\centering
\includegraphics[height=10.3cm]{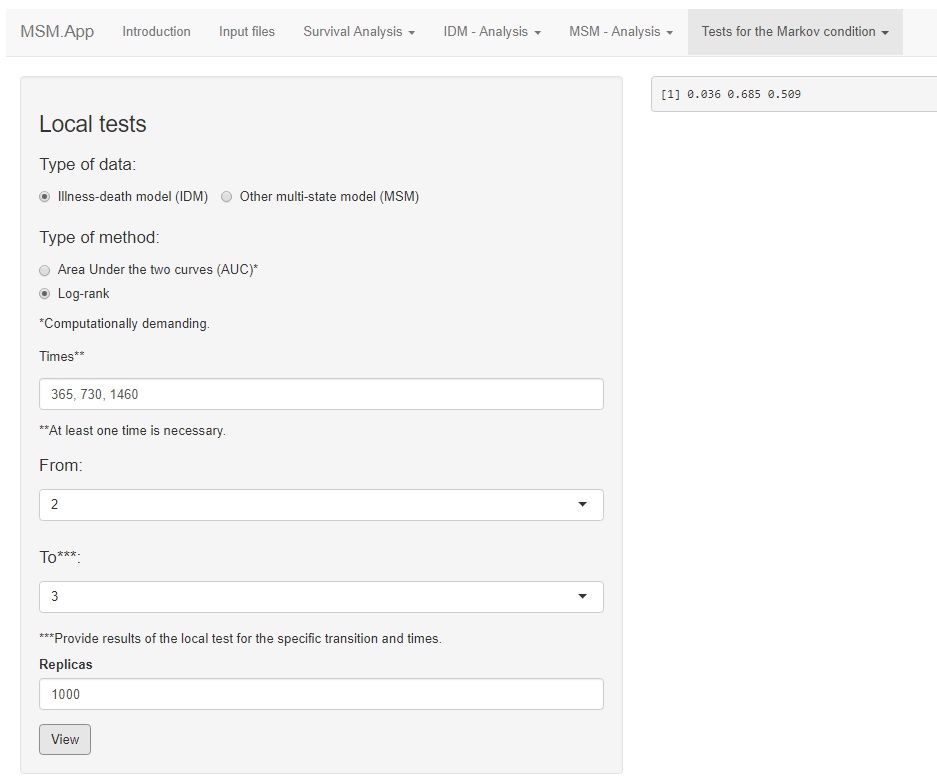}
\caption{Results of the local test for the illness-death model using the colon cancer data set, for \textit{s} = 365, 730 and 1460 days, using the \texttt{Log-rank}  test, for state 2 to state 3.}
\label{fig:local_LR}
\end{figure}

Two types of methods are available for checking the local tests of the Markov condition: (i) the \texttt{AUC} method, which is based on measuring the discrepancy between the \texttt{AJ} estimator of the transition probabilities (which provides consistent estimates when the process is Markovian) and the landmark estimators (which are free of the Markov condition). In this case, the web tool uses the \texttt{LM} estimator for the progressive illness-death models and \texttt{LMAJ} in the case of more complex MSM models; (ii) the \texttt{Log-rank} method, which considers summaries from families of log-rank statistics where patients are grouped by the state occupied at different times. For both types of models (IDM and MSM), the output using the AUC is the same. As an example, in Figure~\ref{fig:markov-localTest}, we obtained the p-values for the local tests based on the $s$ times 365, 730, 1095, 1460, and 1825 for the transition from state 2 to state 3. Results were obtained for an IDM model based on the colon cancer data using 100 replicas. Even though the web form is quite similar to the AUC local test, the output of the log-rank test only provides the results for a specific transition and the times chosen (Figure~\ref{fig:local_LR}).

\vskip 7mm
\textit{Global tests}
\vskip 2mm

\begin{figure} [h]
\centering
\includegraphics[height=6cm]{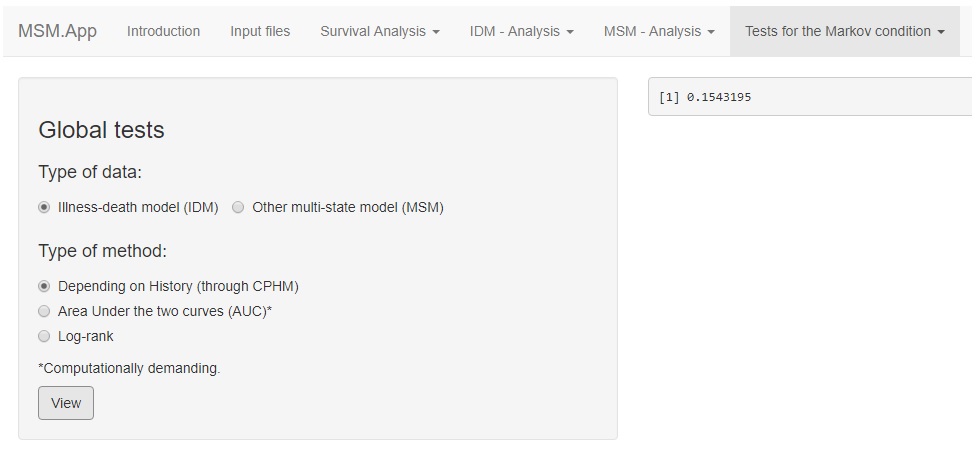}
\caption{Results for this global test given by the Cox PH model to our data indicated that the effect of the time spent in State 1 is not significant (p-value of 0.154), revealing no evidence against the Markov model for the colon data set.}
\label{fig:cox_idm}
\end{figure}

\begin{figure} [h]
\centering
\includegraphics[height=8.5cm]{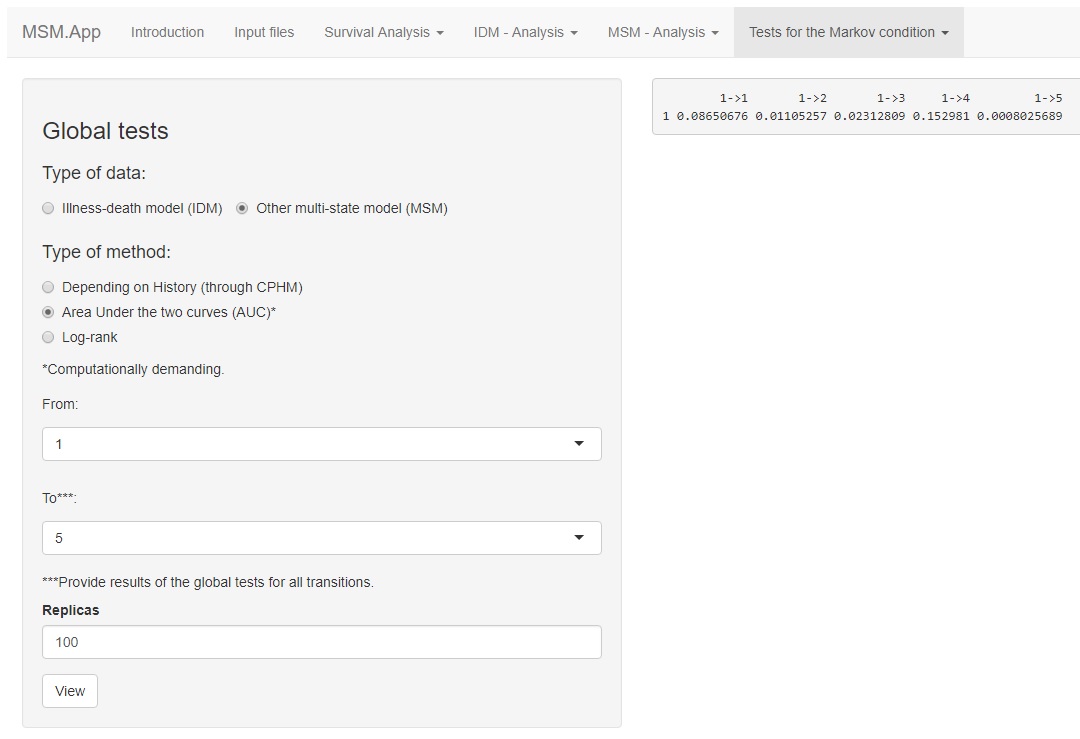}
\caption{Outputs of the global test for the illness-death model based on the \textit{ebmt} data set using the \texttt{AUC} test, from state 1 to state 5. Results for the \texttt{AUC} local test are also shown.}
\label{fig:auc_msm}
\end{figure}

\begin{figure} [h]
\centering
\includegraphics[height=6.7cm]{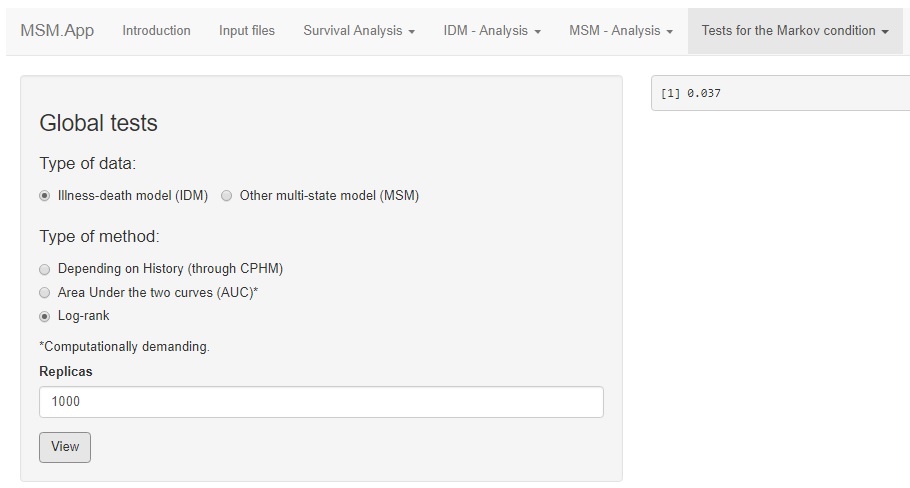}
\caption{Results of the global test for the illness-death model using the colon cancer data set using the \texttt{Log-rank} test.}
\label{fig:global_LR}
\end{figure}

In the \texttt{MSM.app} application, three global tests are available for both IMD and more complex MSM models: (i) the first one is based on Cox models, from which it is possible to evaluate the effect of history on the process. In this case, this can be done by checking the significance of the covariate time until entering the first state of a particular transition. As illustrated in Figure~\ref{fig:cox_idm}, we can conclude that there is no effect of the time spent in the initial state on the transition $2\rightarrow3$ (p-value = 0.1543195), which does not induce the failure of the Markovianity. (ii) the recent proposed global test propose \cite{Soutinho2021c}, based on the area under the curves (\texttt{AUC}), can be used. This test is based on the (\texttt{AUC}) local test results for specific percentiles. The outputs on the right hand of Figure~\ref{fig:auc_msm}  shows the proportion of rejections of the test for all possible transitions between state 1 and state 5. (iii) it is also possible to use the global test based on the log-rank statistics \cite{Titman2020} throughout the similar steps of the previous methods, after selecting Log-rank in the radio button \texttt{HTML} element. The outputs provide only the results of the tests for each transition (Figure~\ref{fig:global_LR}).

\newpage

\noindent \textbf{Acknowledgements}
\vskip 4mm

This research was financed within the research grants PTDC/MAT-STA /28248/2017 and PD/BD/142887/2018.

\vskip 4mm
\baselineskip=12pt


\end{document}